\documentclass[aps,prl,twocolumn,superscriptaddress]{revtex4}
\usepackage{graphicx,amsmath,amssymb,bm}

\newcommand{\kf}{k_{\text{F}}}
\newcommand{\be}{\begin{equation}}
\newcommand{\ee}{\end{equation}}
\newcommand{\beal}{\begin{align}}
\newcommand{\eeal}{\end{align}}
\newcommand{\q}{{\bf q}}
\newcommand{\qp}{{\bf q'}}
\newcommand{\Pt}{{\bf P}}
\newcommand{\pp}{{\bf p}}
\newcommand{\qh}{\widehat{{\bf q}}}
\newcommand{\qph}{\widehat{{\bf q}}'}
\newcommand{\Pth}{\widehat{{\bf P}}}
\newcommand{\la}{\Lambda}
\newcommand{\si}{{\bm \sigma_1}}
\newcommand{\sip}{{\bm \sigma_2}}
\newcommand{\vlk}{$V_{\text{low}\,k}$}

\begin{document}

\title{Polarization contributions to the spin-dependence of the effective
interaction in neutron matter}
\author{Achim Schwenk}
\email[E-mail:~]{aschwenk@mps.ohio-state.edu}
\affiliation{Department of Physics, The Ohio State University,
Columbus, OH 43210, USA}
\author{Bengt Friman}
\email[E-mail:~]{b.friman@gsi.de}
\affiliation{Gesellschaft f\"ur Schwerionenforschung, Planckstr. 1, 64291
Darmstadt, Germany}


\begin{abstract}
We calculate the modification of the effective interaction of
particles on the Fermi surface due to polarization contributions,
with particular attention to spin-dependent forces. In addition to
the standard spin-spin, tensor and spin-orbit forces, spin
non-conserving effective interactions are induced by screening in
the particle-hole channels. Furthermore, a novel long-wavelength
tensor force is generated. We compute the polarization
contributions to second order in the low-momentum interaction
$V_{\text{low}\,k}$ and find that the medium-induced spin-orbit
interaction leads to a reduction of the $^3$P$_2$ pairing gap for
neutrons in the interior of neutron stars.
\end{abstract}


\maketitle

{\it Introduction.} -- Landau-Fermi liquid theory is a powerful
effective theory for strongly interacting Fermi systems at low
temperatures. It has been successfully applied to liquid $^3$He,
nuclear matter and nuclei. While the free interaction between
$^3$He atoms is almost state-independent, the nuclear interaction
is complicated due to large non-central spin-orbit and tensor
forces, which are crucial for understanding nuclear phenomena. For
investigations of matter under extreme conditions, such as nuclei
with large proton or neutron excess and asymmetric nuclear matter
in neutron stars, the role of
non-central forces in the effective interaction must be understood.

As part of a program to determine effective nuclear interactions
using renormalization group methods~\cite{RGnm}, we analyze the
spin-dependence of the quasiparticle interaction and the low-energy
scattering amplitude in the presence of non-central forces. We
focus on pure neutron matter, and as an application, we estime the
modification of the $^3$P$_2$ pairing gap in neutron star interiors
due to the screening of the nucleon-nucleon interaction. This is a
long-standing problem in neutron star structure, and since
polarization effects suppress the S-wave gaps by a factor
four~\cite{RGnm}, large effects may be expected.

{\it Symmetry considerations.} -- In general, the two-body
interaction is hermitian and constrained by
symmetries, specifically, time-reversal and parity
invariance, as well as invariance under exchange of particle
labels. In addition, in a non-relativistic theory in vacuum,
the potential is
Galilean invariant, i.e., independent of the particle-pair momentum
$\Pt = {\bf p}_1 + {\bf p}_2 = {\bf p}_3 + {\bf p}_4$. The possible
operators are well-known~\cite{OM}; scalar: $\openone$, spin-spin:
$\si \cdot \sip$, spin-orbit: $i (\si + \sip) \cdot
\q \times \qp$, tensor: $S_{12}(\q) \equiv \si \cdot \q \, \sip \cdot
\q - 1/3 \, q^2 \, \si \cdot \sip$ (and the exchange thereof
$S_{12}(\qp)$) as well as the quadratic spin-orbit force: 
$\q \cdot \qp \, \bigl( \si \cdot \q \, \sip \cdot \qp + \si \cdot \qp \,
\sip \cdot \q - 2/3 \, \q \cdot \qp \, \si \cdot \sip \bigr)$. The momentum
transfer is $\q = {\bf p}_1 - {\bf p}_3$ and in the exchange term
$\qp = {\bf p}_1 - {\bf p}_4$.

In the many-body medium the presence of the Fermi sea defines a
preferred frame. Therefore, the effective two-body interaction
depends on the two-body center of mass (cm) momentum. This is
physically clear, since the magnitude of the cm momentum, for
given momentum transfers, determines where the interacting
particles are relative to the Fermi sea. For particles on the Fermi
surface $q^2 + q^{\prime\,2} + P^2 = 4 \kf^2$ and the momentum
dependence of possible invariants is constrained geometrically,
since $\q$, $\qp$ and $\Pt$ are orthogonal. As a consequence, the
quadratic spin-orbit force vanishes in this case.

{\it Effective interactions on the Fermi surface.} -- In the
presence of a Fermi sea, additional operators are possible. For
particles on the Fermi surface, these are~\cite{tensordep}
\begin{eqnarray}
&S_{12}(\Pt) & \text{cm tensor} \label{cmtensor} \\
&D_{12}(\q,\Pt) \equiv i (\si - \sip) \cdot \q \times \Pt &
\text{diff vector} \label{diffvector} \\
&A_{12}(\qp,\Pt) \equiv (\si \times \sip) \cdot (\qp \times \Pt) &
\text{cross vector} \label{crossvector}
\end{eqnarray}
Operators $D_{12}$ and $A_{12}$ are related by exchange. These are
antisymmetric in spin and thus do not conserve the spin of the
interacting particle pair. In this Letter, we explore the
microscopic origin of these forces and compute the contributions to
the quasiparticle interaction in neutron matter. Our results can be
used as input for calculations of neutron star properties. For
particles not on the Fermi surface, further invariants are
possible~\cite{offFS}.

\begin{figure}[t!]
\begin{center}
\includegraphics[scale=0.5,clip=]{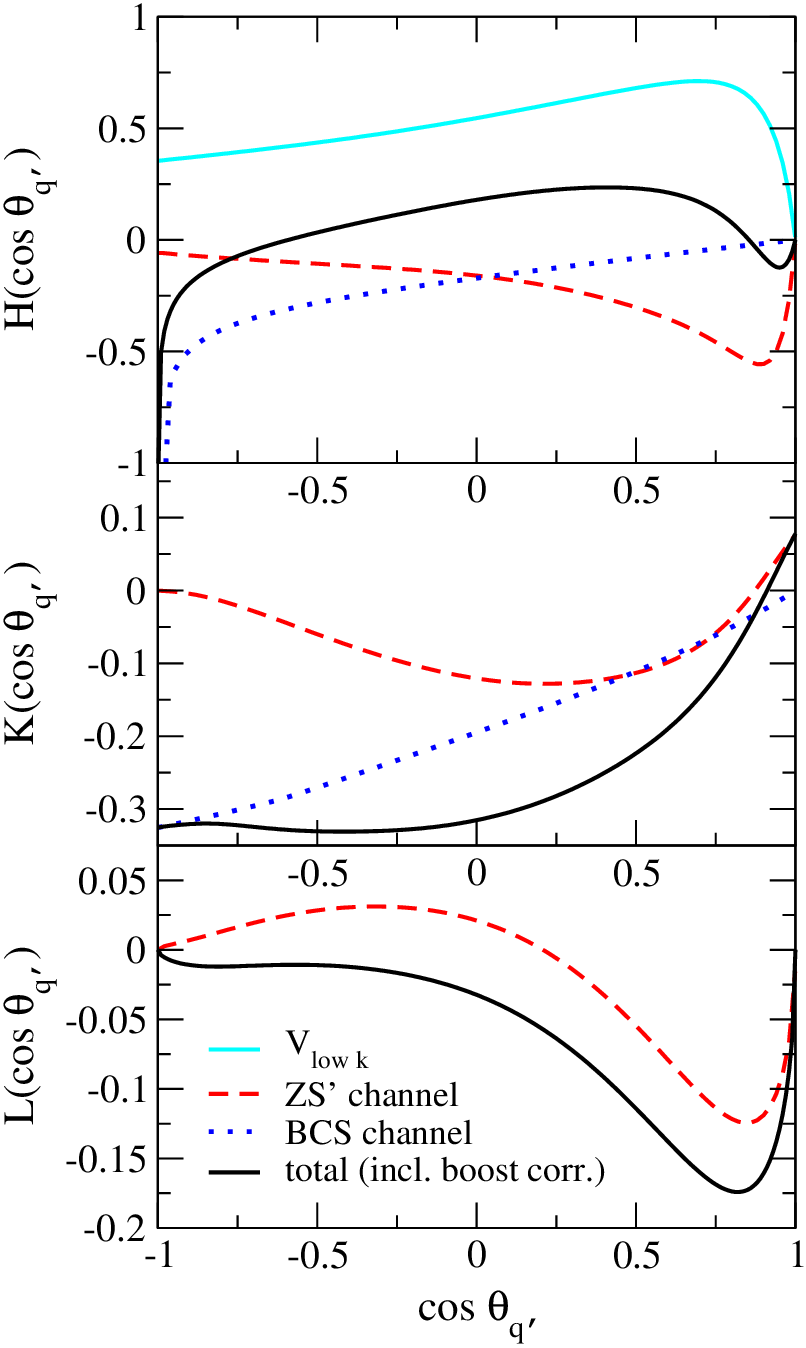}
\end{center}
\caption{Angular dependence ($\cos\theta_\qp=1-q^{\prime\,2}/2 \kf^2$)
of the non-central quasiparticle interactions in neutron matter for
$\kf = 1.7 \, \text{fm}^{-1}$. The $V_{\text{low} k}$ contribution
as well as the polarization effects from the exchange particle-hole
(ZS') and the particle-particle (BCS) channel are shown for
$m^\ast/m = 0.83$.}
\label{noncentral}
\end{figure}

Both the cm tensor and the cross vector operator survive in the
long-wavelength limit, $q \to 0$, and thus contribute to the
quasiparticle interaction in nuclear matter. The exchange tensor in
the quasiparticle interaction was considered in~\cite{Haensel1,Haensel2}, 
and Landau parameters were
computed in~\cite{tensor,Dickhoff}. We introduce the Fermi liquid
parameters, $H_l$, $K_l$, and $L_l$ for the non-central
interactions,
\begin{align}
\mathcal{F}_{\si, \sip}^{\text{n-c}}(\qp,\Pt) &= H(\cos\theta_{\qp})
\, S_{12}(\qph) + K(\cos\theta_{\qp}) \, S_{12}(\Pth) \nonumber \\
&+ L(\cos\theta_{\qp}) \, A_{12}(\qph,\Pth) .
\end{align}
The tensor operators are defined with unit vectors~\cite{OP} and
the dependence on Landau angle $\theta_\qp$ is expanded in Legendre
polynomials, $H(x)=\sum_l H_l \, P_l(x)$ etc. The novel Fermi-liquid
interactions $K$ and $L$ have not been considered in previous work.
Analyticity of the quasiparticle interaction implies $H \sim
q^{\prime\,2}$ as $q'\to 0$, $K \sim P^2$ as $q' \to 2 \kf$ ($P\to
0$) and $L \sim q'$ as $q' \to 0$ (as well as $L \sim P$ as $q'\to
2 \kf$). Many-body effects may give rise to singularities in the
effective interaction, which modify these limits. This is
illustrated by $K(\cos\theta_{\qp})$ in Fig.~\ref{noncentral},
where the pairing singularity cancels the zero of the cm tensor in
the limit $q'\to 2\kf$ ($\cos \theta_{\qp} \to -1$).

{\it Effects of the many-body medium.} -- At second-order, there
are contributions to the effective four-point vertex from
scattering in the BCS channel with intermediate particle-particle
and hole-hole excitations of cm momentum $\Pt$, as well as
scattering in the direct particle-hole or zero sound (ZS) channel
and in the exchange particle-hole (ZS') channel. The ZS and ZS'
channels include intermediate states with particle-hole excitations
of momentum $\q$ and $\qp$ respectively. For the quasiparticle
interaction only the BCS and ZS' channel contribute.
General recoupling arguments imply that the
interference of the spin-spin with the tensor force in the
particle-hole channels leads to a large renormalization of the
tensor interaction~\cite{RGnm}. Moreover, the presence of a third
particle in intermediate states induces novel contributions to the
effective interaction, of the form of Eqs.~(\ref{cmtensor})
- (\ref{crossvector}).

For an antisymmetrized interaction $f^{\text{d}}_{\si, \sip}(\q,\qp)
= V_{\si, \sip}(\q,\qp) - P_{\bm \sigma} V_{\si, \sip}(\qp,\q)$
(where $P_{\bm \sigma}$ is the spin exchange operator and the superscript
$\text{d}$ labels the driving term~\cite{RGnm}), the
particle-hole contributions in the ZS' channel are given by
\begin{align}
&a^{\text{ZS'}}_{\si, \sip}(\q,\qp,\Pt) = - P_{\bm \sigma}
\int \frac{d^3 \pp}{(2 \pi)^3}
\, \frac{n_{\pp + \qp/2} - n_{\pp - \qp/2}}{\varepsilon_{\pp + \qp/2}
- \varepsilon_{\pp - \qp/2}} \nonumber\\
&\times \mathrm{Tr}_{\bm \sigma} \:
f^{\text{d}}_{\si, {\bm \sigma}}(\qp,\frac{\Pt+\q}{2}-\pp) \,
f^{\text{d}}_{{\bm \sigma}, \sip}(\qp,\pp-\frac{\Pt-\q}{2}) ,
\end{align}
where $n_{\bf p}$ denotes the Fermi-Dirac distribution function and
$\varepsilon_{\bf p}$ is the quasiparticle energy for which a
single-particle spectrum with effective mass is used. The exchange
tensor, iterated in the particle-hole ladder, yields an unusual
ordering of the spin operators $\si \cdot {\bf t'} \, \si \cdot
{\bf t} \, \sip \cdot {\bf t} \,\sip \cdot {\bf t'}$, where
${\bf t}^{(\prime)} = \pm (\Pt/2 \pm \q/2 -\pp)$. This ordering
gives rise to a particular coupling between the spin and the
angular motion, which leads to antisymmetric spin operators. Furthermore,
particle-hole polarization contributions involving spin-orbit
forces always result in spin non-conserving interactions.

On the other hand, in the BCS channel the dependence on the cm
momentum enters only through the phase space and the ordering of
the spin operators is the same as in vacuum. Consequently, the BCS
channel does not yield antisymmetric spin operators. The second
order contribution is given by
\begin{align}
&a^{\text{BCS}}_{\si, \sip}(\q,\qp,\Pt) = \frac{1}{2}
\int \frac{d^3 \pp}{(2 \pi)^3} \, \frac{1- n_{\Pt/2+\pp} -
n_{\Pt/2-\pp}}{2 \mu - \varepsilon_{\Pt/2+\pp}
- \varepsilon_{\Pt/2-\pp}} \nonumber\\
&\times f^{\text{d}}_{\si, \sip}(\pp-{\bf k}_{f},\pp+{\bf k}_{f})
\, f^{\text{d}}_{\si, \sip}({\bf k}_{i}-\pp,{\bf k}_{i}+\pp) ,
\end{align}
where ${\bf k}_{i,f} = (\qp \pm \q)/2$ denotes initial/final
relative momenta and $\mu$ is the chemical potential.

\begin{table}
\caption{The $l \leqslant 4$ Fermi liquid parameters for neutron matter
at the Fermi momentum $\kf = 1.7 \, \text{fm}^{-1}$, with effective
mass $m^\ast/m = 0.83$. The different contributions are discussed
in the text. The total includes also the (small) boost corrections.
The renormalization of the quasiparticle strength $z_{\kf}$ is
neglected and we use $z_{\kf}=1$ throughout.}
\label{flparams}
\begin{ruledtabular}
\begin{tabular}{rrrrrr}
Landau $l$  & $0$  & $1$ & $2$  & $3$  & $4$ \\ \hline\hline
\vlk \\ \hline
scalar $F_l^{\text{d}}$
& $-0.734$ & $-0.498$ & $-0.200$ & $-0.068$ & $-0.052$ \\
spin-spin $G_l^{\text{d}}$
& $0.842$ & $0.412$ & $0.219$ & $0.109$ & $0.053$ \\
exch. tensor $H_l^{\text{d}}$
& $0.529$ & $0.150$ & $-0.096$ & $-0.141$ & $-0.124$
\\ \hline\hline
ZS' channel \\ \hline
scalar 
& $0.552$ & $0.406$ & $0.119$ & $0.131$ & $0.099$ \\
spin-spin 
& $0.024$ & $-0.038$ & $-0.052$ & $-0.016$ & $0.002$ \\
exch. tensor 
& $-0.214$ & $-0.218$ & $-0.086$ & $0.004$ & $0.067$ \\
cm tensor 
& $-0.071$ & $-0.014$ & $0.104$ & $0.047$ & $0.009$ \\
cross vector 
& $-0.015$ & $-0.073$ & $-0.057$ & $0.018$ & $0.032$
\\ \hline\hline
BCS channel \\\hline
scalar 
& $-0.291$ & $0.187$ & $-0.146$ & $0.121$ & $-0.088$ \\
spin-spin 
& $0.032$ & $0.126$ & $-0.006$ & $0.032$ & $-0.019$ \\
exch. tensor 
& $-0.201$ & $0.249$ & $-0.089$ & $0.073$ & $-0.060$ \\
cm tensor 
& $-0.187$ & $0.160$ & $0.020$ & $0.002$ & $0.005$
\\ \hline\hline
total \\\hline
exch. tensor $H_l$
& $0.077$ & $0.140$ & $-0.266$ & $-0.047$ & $-0.101$ \\
cm tensor $K_l$
& $-0.258$ & $0.147$ & $0.124$ & $0.048$ & $0.014$ \\
cross vector $L_l$
& $-0.061$ & $-0.089$ & $-0.035$ & $0.025$ & $0.041$ \\
\end{tabular}
\end{ruledtabular}
\end{table}

{\it Boost corrections.} -- In addition to the
dynamical effects, there are kinematical contributions from
boosting the two-body interaction to the rest frame of the Fermi
sea. To leading order in $\kf^2/m^2$, the boost to the frame where
the nucleon pair carries momentum $\Pt$ is given in terms
of the vacuum interaction in the cm frame~\cite{FF,Forest}
\begin{align}
\delta &\mathcal{F}_{\si, \sip}(\q,\qp,\Pt) = - \frac{P^2}{4 \, m^2} \,
\mathcal{F}^{\text{d}}_{\si, \sip}(\q,\qp) \nonumber \\
&+ \frac{i}{16 \, m^2} \, (\si - \sip) \times \Pt \cdot (\qp - \q) \,
\mathcal{F}^{\text{d}}_{\si, \sip}(\q,\qp) \nonumber \\
&- \frac{i}{16 \, m^2} \, \mathcal{F}^{\text{d}}_{\si, \sip}
(\q,\qp) \, (\si - \sip) \times \Pt \cdot (\qp + \q) ,
\label{relcorr}
\end{align}
where $\mathcal{F}^{\text{d}}= m^\ast \kf f^{\text{d}}/\pi^2$. In
Eq.~(\ref{relcorr}) the particles are restricted to the
Fermi surface and direct and exchange terms are included.
In the long-wavelength limit, we find
\begin{align}
\delta &\mathcal{F}_{\si, \sip}(\qp,\Pt) = - \frac{P^2}{4 \, m^2} \,
\bigl\{ F^{\text{d}}(\cos\theta_{\qp}) \nonumber \\
&+ G^{\text{d}}(\cos\theta_{\qp}) \, \si \cdot \sip +
H^{\text{d}}(\cos\theta_{\qp}) \, S_{12}(\qph) \bigr\} \nonumber \\
&- \frac{1}{4 \, m^2} \, A_{12}(\qp,\Pt) \bigl\{ G^{\text{d}}
(\cos\theta_{\qp}) + \frac{1}{6} \, H^{\text{d}}(\cos\theta_{\qp}) \bigr\} ,
\label{reqpint}
\end{align}
with standard notation for the scalar and spin-spin parts,
$F^{\text{d}}$ and $G^{\text{d}}$. We note that the boost
corrections contribute to the spin non-conserving part, but not to
the cm tensor Eq.~(\ref{cmtensor}). These kinematical effects are
of order $\kf^2/m^2$, while the many-body effects are of order
$(m^\ast/m) \, \kf \langle V \rangle$. Here, the brackets denote
the angular average over the interaction in the loop integral.

{\it Results.} -- We start from the free-space
low-momentum interaction $V_{\text{low} k}$~\cite{Vlowk} with a
density-dependent cutoff $\la = \sqrt{2} \, \kf$~\cite{RGnm}.
Detailed results for both neutron and nuclear matter will be
presented elsewhere.
Here, we focus on the novel
spin-dependent interactions. The contributions to the Fermi liquid
parameters are given in Table~\ref{flparams} and the dependence on
$\cos\theta_\qp$ is shown in Fig.~\ref{noncentral}. We find a
substantial renormalization of the exchange tensor, which
necessitates a self-consistent treatment within, e.g., the RG
approach~\cite{RGnm}, and significant contributions to the new
interaction terms. In particular, the cm tensor is comparable to
the exchange tensor at this order.

\begin{figure}[t!]
\begin{center}
\includegraphics[scale=0.5,clip=]{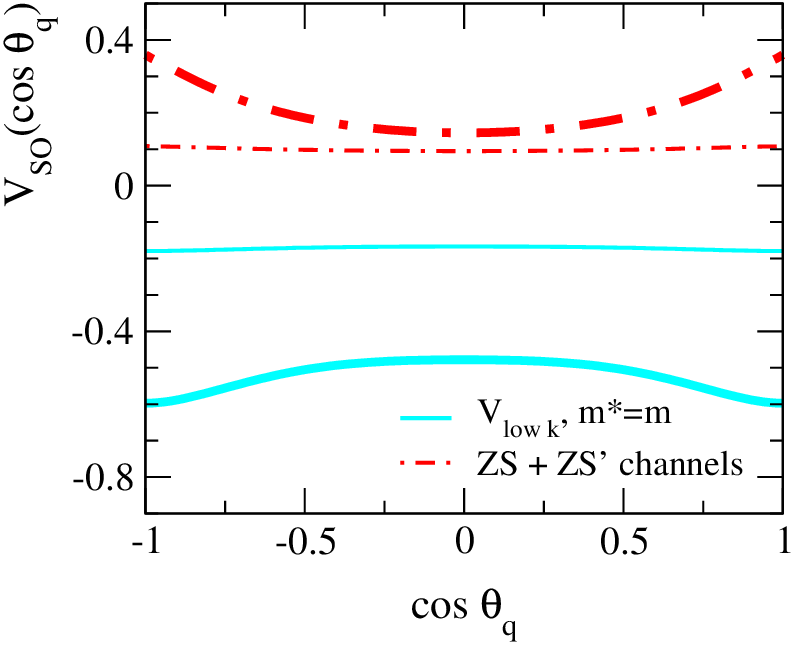}
\end{center}
\caption{Comparison of the particle-hole induced spin-orbit amplitude to
the free-space interaction. Results are presented for back-to-back
scattering, $P=0$, $\kf = 1.7 \, \text{fm}^{-1}$ (thick lines) and
$\kf = 1.0\, \text{fm}^{-1}$ (thin lines). Here, the
(antisymmetrized) spin-orbit force is shown in units of the density
of states with the operator $V_{\text{SO}} (\cos\theta_\q)
\, i (\si + \sip) \cdot \q/\kf \times \qp/\kf$.}
\label{soinduced}
\end{figure}

In calculations of transport processes and pairing phenomena, one
needs the scattering amplitude at finite momentum, $q
\neq 0$. In particular, the in-medium modification of the
spin-orbit interaction is of special interest, because it is
crucial for an accurate reproduction of the P-waves in free
space, and similarly for a realistic assessment of P-wave pairing
in neutron stars. As shown in Fig.~\ref{soinduced}, we find a
repulsive induced spin-orbit interaction
due to particle-hole screening. In order to qualitatively
understand the resulting interaction, we assume a
contact spin-spin and an averaged spin-orbit matrix element. Then,
the ZS and ZS' channel for $P=0$ are repulsive
\begin{equation}
V_{\text{SO}}^{\text{ind}}(\cos\theta_\q) = - G_0^{\text{d}}
\, \frac{m^\ast}{2 \, m}
\, \langle V_{\text{SO}}^{\text{d}}
\rangle \, \bigl( U(q/\kf) + U(q'/\kf) \bigr) ,
\label{simplearg}
\end{equation}
for an attractive spin-orbit force. Here $U(q/\kf)$ denotes the
(positive) static Lindhard function and $q^{(\prime)} =
\kf \sqrt{2 \mp 2\cos\theta_\q}$. As in Fig.~\ref{soinduced},
$V_{\text{SO}}^{\text{ind}}$ is only weakly dependent on
$\cos\theta_\q$. The contributions due to the mixing of spin-orbit
and tensor forces are also repulsive, with a similar but more
complicated momentum dependence.

{\it Triplet pairing.} -- To illustrate the importance of
non-central induced interactions, we estimate the angle-averaged
gap using weak coupling BCS theory~\cite{PZ} (for details
see~\cite{Cugnon,Baldo}). The coupling to the $^3$F$_2$ partial
wave is neglected in this exploratory calculation. In the weak
coupling approximation, the $^3$P$_2$ gap is then given by
\begin{equation}
\Delta _{^3\text{P}_2}
= \frac{\kf^2}{m} \, \exp\biggl( \frac{\pi}{2 \, \kf \, m \,
V_{\text{low}\,k; \text{$^3$P}_2}(\kf,\kf)} \biggr) ,
\label{3p2weak}
\end{equation}
where the arguments of $V_{\text{low} k}$ are the magnitude of the
relative momenta $|{\bf k}_{i,f}| = \kf$. When particle-hole screening
effects are included, the pairing interaction $V_{\text{low}\,k;
\text{$^3$P}_2}(\kf,\kf)$ is replaced by the $^3$P$_2$ projection of
the effective interaction in the particle-particle channel. The
high-lying states in the BCS channel are included in $V_{\text{low}
k}$ while the low-lying part is accounted for through the
(approximate) solution of the gap equation.
\begin{figure}[t!]
\begin{center}
\includegraphics[scale=0.45,clip=]{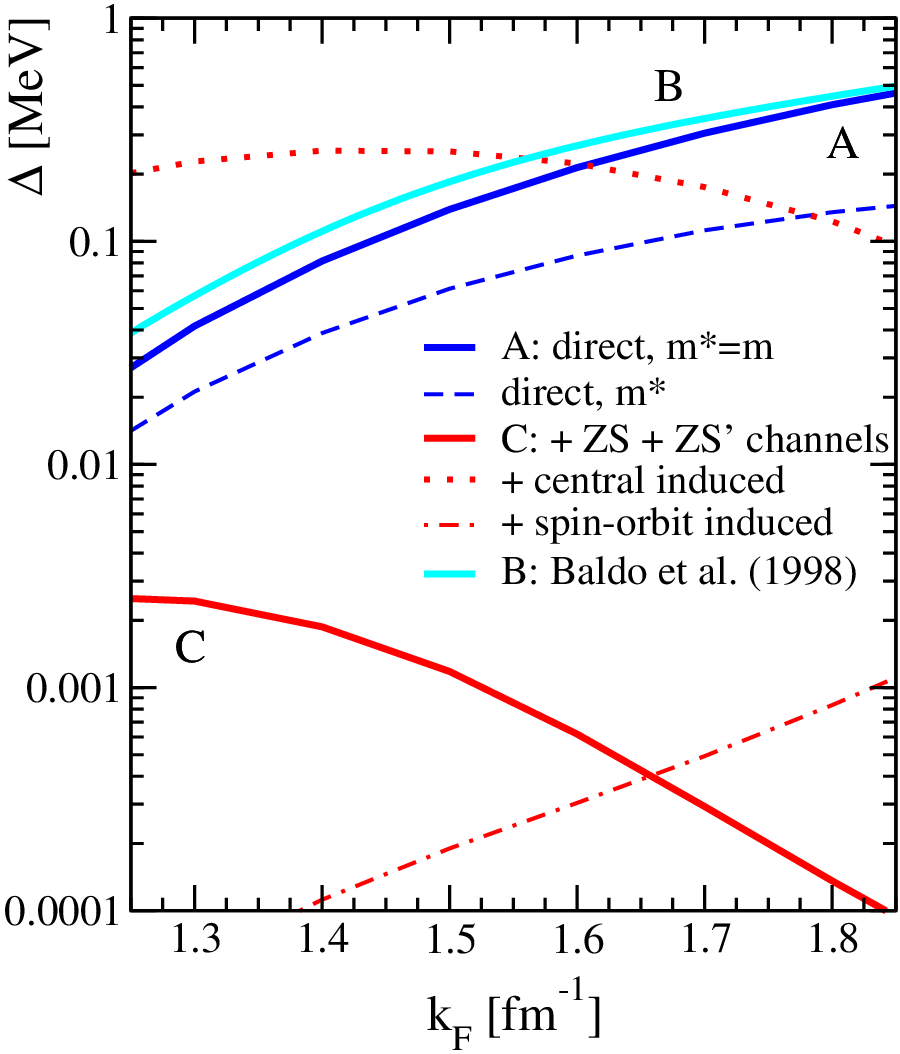}
\end{center}
\caption{The angle-averaged pairing gap $\Delta _{^3\text{P}_2}$ in the
$^3$P$_2$ channel versus the Fermi momentum in neutron matter. The
direct ($V_{\text{low} k}$) pairing gap, computed with the free and
with an effective neutron mass, as well as the gap including
particle-hole polarization effects on the pairing interaction are
shown. We also give the gap, obtained when only the central and
only the spin-orbit polarization contributions are taken into
account. For reference, we show the results of Baldo {\it et
al.}~\cite{Baldo}, obtained by solving the BCS gap equation in the
coupled $^3$P$_2$--$^3$F$_2$ channel for different free-space
interactions. $V_{\text{low} k}$ is obtained from the CD Bonn
potential.}
\label{3p2gaps}
\end{figure}

We find that spin-dependent polarization effects reduce the
$^3$P$_2$ pairing gap (C compared to A and B in
Fig.~\ref{3p2gaps}). This is in contrast to the increase of 
the $^3$P$_2$ gap, which one obtains when polarization effects due only
to central forces are included (dotted vs. dashed line in
Fig.~\ref{3p2gaps})~\cite{3pf2}. The reduction of the gap is
predominantly due to the repulsion from the medium-induced
spin-orbit force, discussed above. This effect was not taken into
account in earlier work. Note that for the densities given in
Fig.~\ref{3p2gaps}, the ratio of the induced pairing matrix element
to the free-space $V_{\text{low} k}$ contribution ranges from
$0.15-0.5$. The significant reduction of the gap for only moderate
changes of the pairing interaction is due to the singular
dependence on the matrix element in Eq.~(\ref{3p2weak}) for small gaps. 
We emphasize that our results are qualitative; a quantitative
calculation should include the full polarization contributions for
non-central interactions and the solution of the full coupled-channel
BCS gap equation. We present results only for $\kf \lesssim 2\, 
\text{fm}^{-1}$, where the NN interaction is strongly constrained 
by data.

{\it Conclusions.} -- In summary, we have found novel non-central
effective nuclear interactions, with spin non-conserving forces
induced by particle-hole polarization effects. In microscopic shell
model calculations, these are implicitly included in the
polarization force of Kuo and Brown~\cite{KuoBrown}. Furthermore,
the renormalization of the spin-orbit interaction in the medium has
important consequences for P-wave pairing. The resulting
suppression of the superfluid gap has direct impact on the
properties of neutron stars and on their cooling by neutrino
emission~\cite{cooling,Paulo}. The implications of the new
interactions for nuclear spectra, for the spin-isospin response and
neutrino transport in stellar collapse, for magnetic
susceptibilities (see~\cite{OP}), for deformations of the Fermi
surface in spin-polarized systems, and for spin relaxation and
mixing of spin and density waves remain to be investigated. Since
the new interactions contribute only to spin non-conserving 
transitions, it may be possible to observe these effects in 
scattering with polarized beams.

We are grateful to Gerry Brown, Dick Furnstahl, Emma Olsson, Chris Pethick
and Dan-Olof Riska for useful discussions. The work of AS is supported by
the NSF under Grant No. PHY-0098645 and PHY-0139973.


\begin{thebibliography}{99}
\bibitem{tensordep} The given tensor operators are linearly
dependent, i.e., $S_{12}(\qh) + S_{12}(\qph) + S_{12}(\Pth)=0$.
In this work, we have kept all tensor operators
explicitly and eliminate $S_{12}(\qh)$ for long-wavelength and
$S_{12}(\Pth)$ for pairing interactions.
\bibitem{offFS} These are a spin-orbit
like $\qp\cdot \Pt \, i (\si + \sip) \cdot \q \times \Pt$, two more
difference vectors, $\qp \cdot \Pt \, D_{12}(\q,\qp)$ and $\q
\cdot \qp \, D_{12}(\qp,\Pt)$, and finally $\q \cdot \Pt \,
S_{12}(\q,\Pt)$ as well as the corresponding operators obtained by
exchange. The last operator has the same spin-momentum structure as
the quadratic spin-orbit with $\qp$ replaced by $\Pt$.
\end{thebibliography}
\end{document}